\documentclass[prl, aps, nofootinbib, twocolumn, amssymb, superscriptaddress, 10pt]{revtex4-2}
\usepackage{amsmath}
\usepackage{amssymb}
\usepackage{amsthm}
\usepackage{amsfonts}
\usepackage{enumerate}
\usepackage{latexsym}
\usepackage{color}
\usepackage{xcolor}
\usepackage{setspace} 
\usepackage{blindtext}
\usepackage{dsfont}
\usepackage{mathrsfs}
\usepackage[normalem]{ulem}

\usepackage{lipsum}

\usepackage{tabularx}
\newcolumntype{Y}{>{\centering\arraybackslash}X}

\usepackage{stackengine}
\usepackage{bbold}

\usepackage{bm}
\usepackage{graphicx}

\usepackage{hyperref}
\hypersetup{
pdfnewwindow=true, colorlinks=true,
linkcolor=blue, anchorcolor=blue,
citecolor=blue, filecolor=blue,
menucolor=blue, urlcolor=blue} 

\usepackage{pifont}
\newcommand\scalemath[2]{\scalebox{#1}{\mbox{\ensuremath{\displaystyle #2}}}}

\newcommand{\beginsupplement}{
        \setcounter{table}{0}
        \renewcommand{\thetable}{S\arabic{table}}
        \setcounter{figure}{0}
        \renewcommand{\thefigure}{S\arabic{figure}}
        \setcounter{equation}{0}
        \renewcommand{\theequation}{S\arabic{equation}}
        \setcounter{section}{0}
        \renewcommand{\thesection}{\Alph{section}}
        \setcounter{subsection}{0}
        \renewcommand{\thesubsection}{\arabic{subsection}}
        \setcounter{subsubsection}{0}
        \renewcommand{\thesubsubsection}{\alph{subsubsection}}
}

\newcommand{\vk}{{\mathbf{k}}}

\newcommand{\vq}{{\mathbf{q}}}

\begin{document}

\title{
Kondo-lattice phenomenology of twisted bilayer \texorpdfstring{WSe$_2$}{WSe2} from compact molecular orbitals of topological bands
}

\author{Fang Xie}
\thanks{\href{fx7@rice.edu}{fx7@rice.edu}}
\affiliation{Department of Physics \& Astronomy,  Extreme Quantum Materials Alliance, Smalley-Curl Institute, Rice University, Houston, Texas 77005, USA}
\affiliation{Rice Academy of Fellows, Rice University, Houston, Texas 77005, USA}

\author{Chenyuan Li}
\affiliation{Department of Physics \& Astronomy,  Extreme Quantum Materials Alliance, Smalley-Curl Institute,
Rice University, Houston, Texas 77005, USA}
\affiliation{Rice Academy of Fellows, Rice University, Houston, Texas 77005, USA}

\author{Jennifer Cano}
\affiliation{Department of Physics and Astronomy, Stony Brook University, Stony Brook, New York 11794, USA}
\affiliation{Center for Computational Quantum Physics, Flatiron Institute, New York, New York 10010, USA}

\author{Qimiao Si}
\thanks{\href{qmsi@rice.edu}{qmsi@rice.edu}}
\affiliation{Department of Physics \& Astronomy,  Extreme Quantum Materials Alliance, Smalley-Curl Institute, Rice University, Houston, Texas 77005, USA}

\date{\today}

\begin{abstract}
    The discovery of superconductivity and correlated electronic phases in twisted bilayer WSe$_2$ (\href{https://doi.org/10.1038/s41586-024-08116-2}{Xia et al., Nature 2024}; \href{https://doi.org/10.1038/s41586-024-08381-1}{Guo et al., Nature 2025}) has generated considerable excitement. 
    Accompanying the superconductivity and a correlated insulator phase is the Kondo-lattice-like phenomenology in transport properties.
    Here we consider how such phenomenology can develop when the combination of the active bands are topological. 
    We advance a unique construction of compact molecular orbitals through a partial Wannierization that is symmetry preserving.
    The resulting Anderson lattice model provides the basis for a microscopic understanding of the experimental observation, including the involved energy scales. Our approach may apply to a broad range of settings where topology and correlations interplay.
\end{abstract}

\maketitle

{\it \color{blue} Introduction}---
Twisted bilayer transition metal dichalcogenides (TMDC) have recently gained significant attention as a platform for exploring strongly correlated quantum phases, including Mott insulators \cite{Wang2020Correlated}, heavy fermion metals \cite{zhao2022gate}, and superconductors \cite{Xia2024Unconventional,Guo2024Superconductivity}. 
Such correlated phenomena bear striking analogies with their counterparts of bulk quantum materials \cite{Keimer2017,Lee-RMP06,Pas21.1}.
The moir\'e potential created by the relative twisting of two monolayers introduces flat electronic bands, where Coulomb interactions dominate over kinetic energy, leading to emergent many-body effects. 
While superconductivity has been widely studied in twisted bilayer graphene, its discovery in twisted WSe$_2$ has sparked intense interest.
Various theoretical models have been proposed to understand the superconductivity and related correlation physics in this system \cite{Xie2024Superconductivity,Schrade2024Nematic,Kim2025,Guerci2024Topological,Chen2023Singlet,Zegrodnik2023Mixed,Zhou2023Chiral,Christos2024Approximate,Crepel2024bridging, Tuo2024Theory,Fischer2024Theory, Chubukov2025Quantum, Zhu2025Superconductivity,MunozSegovia2025Twist}.
To make progress, it is worth noting that superconductivity develops near correlated phases and, moreover, the superconducting transition temperature $T_c$ reaches as high as a few percent of the effective Fermi temperature; both features suggest that the observed superconductivity is unconventional.

Depending on the carrier concentration and displacement field strength, a correlated insulator phase anchors the development of superconductivity \cite{Xia2024Unconventional}.
It shows the Kondo-lattice-like phenomenology in transport properties \cite{Hu-Natphys2024,Kirchner_RMP}: the resistivity showing a characteristic peak in its temperature dependence, signifying the onset of Kondo coherence, and the resistivity at the peak temperature corresponds to a mean free path that is on the order of the Fermi wavelength. 
Importantly, the involved bands are expected to be topological. 
In particular, the top most moir\'e bands in twisted bilayer WSe$_2$ at twisting angle $\theta = 3.65^\circ$ carry nonzero valley Chern number.
Accordingly, understanding the Kondo-lattice-like phenomenology not only paves the way for the development of the superconducting state but also is of inherent interest as a novel correlation phenomenon in topological settings.
A key challenge lies in the topological obstruction to constructing maximally localized Wannier functions for the low-energy moir\'e bands. 

In this work, we overcome this topological obstruction by developing a ``{\it partial Wannierization}''  approach that describes the top two moir\'e bands with nonzero total valley Chern numbers.
This yields a hybrid two-orbital description: one orbital is a maximally localized Wannier function (MLWF) that captures most of the spectral weight of the topmost band, while the other is a topological power-law orbital (TPLO) reflecting the band's nontrivial topology. 
The construction respects all symmmetries of the system and enables a generalized Hubbard model formulation that accurately captures the band geometry and interaction effects. 
We find that the MLWF is close to half-filling and, thus, hosts the dominant effect of strong electronic correlations. 
This framework provides a microscopic and symmetry-respecting platform to explore the interplay between topology and strong correlations in twisted TMDCs, and represents a new paradigm that can be applied to a broad range of correlated topological systems.

\begin{figure}[t]
    \centering
    \includegraphics[width=\linewidth]{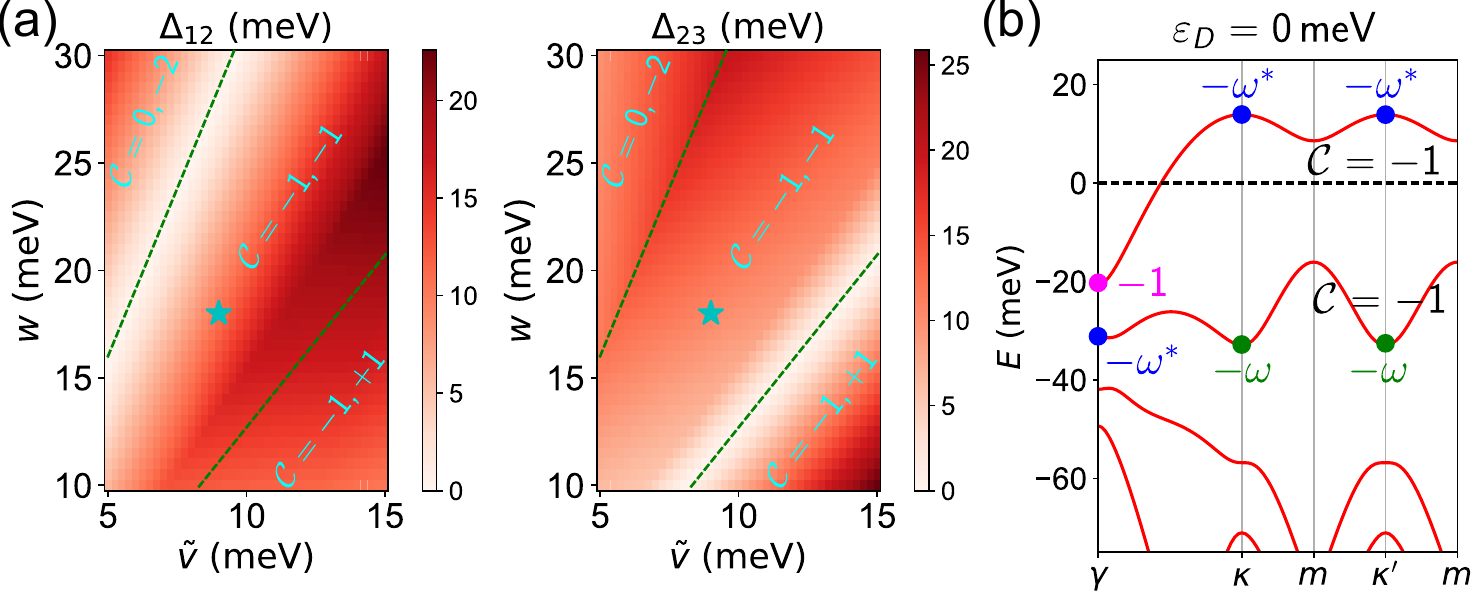}
    \caption{(a) The nature of the active bands' topology in the continuum model at twisting angle $\theta = 3.65^\circ$.
    Here $\tilde{v}$ ($w$) stands for the strength of the intralayer (interlayer)
    moir\'e potential.
    The color coding represents the band gap between the top two moir\'e bands $\Delta_{12}$, and the band gap between the second and the third moir\'e bands $\Delta_{23}$.
    Green dashed lines indicate band gap closing and topological phase transition.
    The Chern numbers of the top two bands are also labeled.
    (b) Single valley band structure of twisted bilayer WSe$_2$ at twisting angle $\theta = 3.65^\circ$.
    The spinful $C_{3z}$ eigenvalues at high symmetry points $\gamma$, $\kappa$ and $\kappa'$ are labeled.
    Here we use the moir\'e potential strength marked by the star symbol in (a), and $\omega = e^{i\frac{2\pi}{3}}$.
    }
    \label{fig:tmd_cm}
\end{figure}

\begin{figure*}[t]
    \centering
    \includegraphics[width=\linewidth]{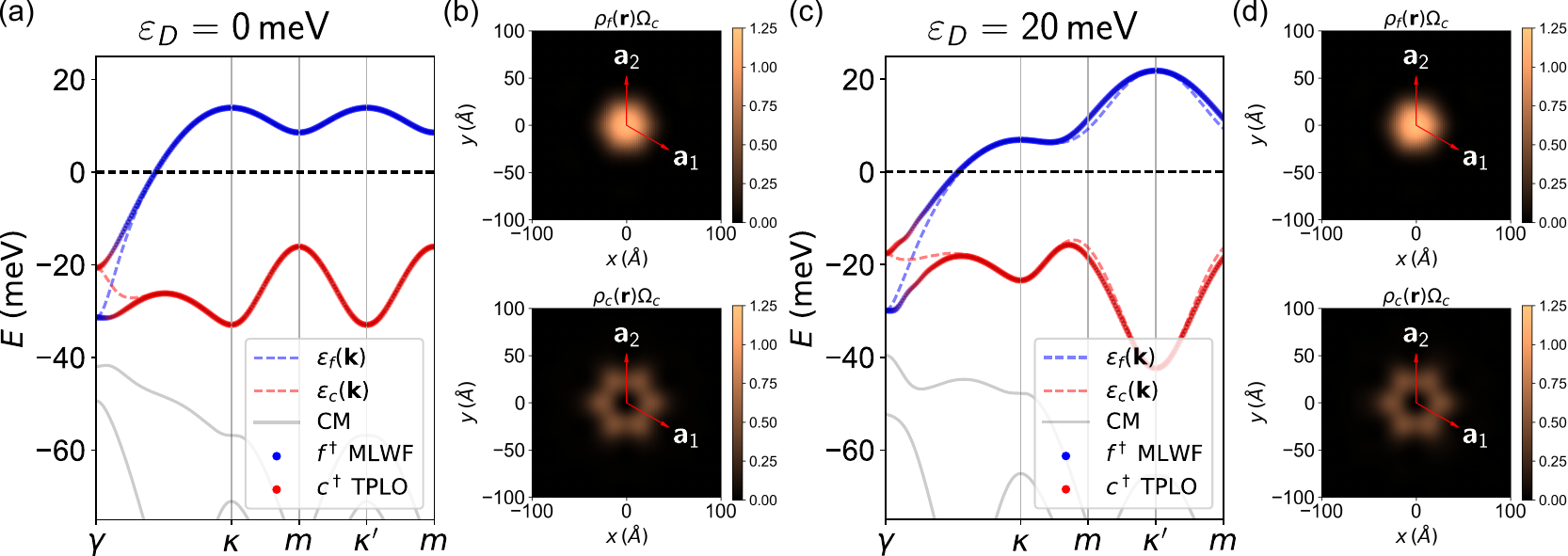}
    \caption{(a) The band structure and orbital projections at displacement field $\varepsilon_D = 0 \rm\,meV$.
    Here, the grey lines stand for the band structure of the continuum model. 
    Blue and red markers stand for the orbital contents of the localized $f$ orbital (maximally localized Wannier function, MLWF) and the conduction $c$ band (topological power-law orbital, TPLO).
    The blue and red dashed lines stand for the energies of the MLWF and TPLO bands without the hybridization between them.
    (b) The charge density distributions of the Wannier functions of the localized $f$ orbital (upper panel) which decays exponentially, and the topological conduction $c$ band (lower panel) which decays as $1/r^2$.
    Red arrows stand for the basis vectors of the moir\'e superlattice.
    (c) The band structure and orbital projections with displacement field potential strength $\varepsilon_D = 20 \rm\,meV$.
    (d) The charge density distributions of the Wannier functions with displacement field potential strength $\varepsilon_D = 20\,\rm meV$.
    All of the figures are calculated at twisting angle $\theta = 3.65^\circ$.
    }
    \label{fig:disp-field-band}
\end{figure*}

{\it \color{blue} Compact molecular orbitals of topological bands: Partial Wannierization}---
The strong spin-orbit coupling in single-layer TMDC materials 
locks the spin and valley degrees of freedom together \cite{Liu2013Three,Kormanyos2015kp}.
Therefore, the low-energy degrees of freedom can be well-captured by quadratic hole pockets near the $K$ and $K'$ points in the single-layer Brillouin zones.
When the two layers are stacked on top of each other, electronic states with the same spin orientation will hybridize with each other, and a small twisting angle will lead to a moir\'e superlattice, corresponding to the moir\'e Brillouin zone (MBZ).
The twisted bilayer system will inherit the $C_{3z}$ symmetry from the monolayer. 
We describe the band structure in terms of a continuum model~\cite{Wu2019topological,Devakul2021Magic}, which is outlined in the supplemental material (SM; Sec.~A) \cite{supplemental_material}.
The topology of the low-energy moir\'e bands is sensitive to the choice of the inter-layer ($w$) and intra-layer ($\tilde{v}$) moir\'e potentials.
As shown in Fig.~\ref{fig:tmd_cm}(a), moderate changes to these parameters could lead to different Chern numbers of the top bands.
In this paper, we work with the moir\'{e}-potential parameters \cite{Devakul2021Magic} that are labeled by the star symbol in Fig.~\ref{fig:tmd_cm}(a), where the top two moir\'e bands carry the same Chern number $\mathcal{C} = - 1$.
This is in contrast to Ref.~\cite{Xie2024Superconductivity}, which considered a different regime of the potentials such that the top two moir\'e bands carry the opposite Chern numbers.
The corresponding band structure of tWSe$_2$ at twisting angle $\theta = 3.65^\circ$ and zero displacement field is shown in Fig.~\ref{fig:tmd_cm}(b).
The colored characters highlighted in this figure are the spinful $C_{3z}$ eigenvalues of the Bloch states at the three high symmetry points $\gamma$, $\kappa$ and $\kappa'$.

Since the top two bands of such moir\'e structure carry the same non-zero Chern number, they cannot be symmetrically Wannierized into exponentially localized orbitals.
We first note that, the spinful $C_{3z}$ eigenvalues of the top most band at $\kappa$ and $\kappa'$ are both $-\omega^*$, and at $\gamma$ is $-1$.
This means the top most band does not form an elementary representation (EBR) of the space group $P3$ (no.~143) \cite{Cano2021band}.
However, an ``inversion'' of the second top band at $\gamma$ point allows the $C_{3z}$ eigenvalues of the combined wave function to be $-\omega^*$ at all high symmetry points, which corresponds to the EBR induced by the ${}^2\overline{E}$ representation of the $1a$ Wyckoff position ($\mathbf{r}_{1a} = \mathbf{0}$).
This induced representation is usually denoted as $({}^2\overline{E})_{1a} \uparrow P3$.
An exponentially localized Wannier function, which predominantly overlaps with the top most band except for the region around the $\gamma$ point, could then be realized.

Thus, our key idea is to construct a two-orbital model, with one of the orbitals being a maximally localized Wannier function (MLWF), and another orbital being a topological power-law orbital (TPLO), which corresponds to a Chern band with $\mathcal{C} = -2$ \cite{thouless_wannier_1984}. 
The size of MLWF will be comparable to the moir\'e unit cell, which is much larger than that of individual atoms. 
As such, the MLWF can be considered as an effective ``molecular orbital'', and yet it still is compact on account of being less extended than its orthogonal counterpart.
Due to the ``band-inversion'' at the $\gamma$ point, the TPLO and the MLWF will also hybridize with each other.
Our procedure draws some analogy with the construction of the compact molecular orbitals in kagome and related frustrated-lattice  systems whose topological indices add up to zero \cite{hu2023coupled,chen2024emergent, chen2023metallic}, with, however, a crucial difference: in our case the orthogonal orbital is not exponentially localized but instead has a power-law decay. 
It also draws inspiration from the ``reduced Wannier representation'' in the case of a single Chern band~\cite{Cole2024Reduced,Monsen2024Supercell}, though, importantly, our construction preserves all the symmetries of the Hamiltonian.

We perform a disentanglement Wannierization procedure provided by \textsc{Wannier90} \cite{Marzari1997Maximally,Souza2001Maximally,Pizzi2020wannier}.
The wave function of the MLWF (which will be denoted as $f$ orbital) is constructed from a {\it globally smooth} $\vk$-dependent linear combination of Bloch states from the top two moir\'e bands, with the transformation parameters provided as output by \textsc{Wannier90}.
The wave functions of the remaining TPLO ($c$ orbital) can be consequently constructed via a simple orthogonalization process.
Since a globally smooth gauge for the Bloch states of TPLO is prevented by topological obstruction, we fixed it using the algorithm introduced in Ref.~\cite{Xie2024Chern}, placing a vortex with vorticity $\mathcal{C} = -2$ at the $\gamma$ point.

Using the wave functions of the MLWF and TPLO, we can also compute the hopping and hybridization amplitudes among the $f$ and $c$ orbitals in different unit cells.
Therefore, an effective Hamiltonian that captures the subspace of the top two bands can be written as:
\begin{align}
    H_0 =& \sum_{\vk,\sigma}\varepsilon_f^{(\sigma)}(\vk) f^\dagger_{\vk\sigma} f_{\vk\sigma}\ +\sum_{\vk, \sigma} \varepsilon^{(\sigma)}_c(\vk) c^\dagger_{\vk\sigma} c_{\vk\sigma} \nonumber\\
    & + \sum_{\vk, \sigma} \left(V^{(\sigma)}_{\rm hyb}(\vk) f^\dagger_{\vk\sigma} c_{\vk\sigma} + {\rm h.c.}\right),\label{eqn:eff-kin}
\end{align}
in which the parameters for the spin $\uparrow$ and spin $\downarrow$ sectors are related via time-reversal transformation.
Numerical calculation shows that the hopping between these $f$ orbitals is primarily dominated by nearest-neighbor hopping $|t_f| \approx 3.5\,\rm meV$.
Hence, the dispersion of the MLWF can be well approximated by the following form:
\begin{equation}
    \varepsilon^{(\sigma)}_f(\vk) \approx 2|t_f| \sum_{i=1}^3 \cos(\vk \cdot \bm{\delta}_i + \sigma\phi_f) + \tilde{\epsilon}_f\,,
\end{equation}
where the Bravais lattice vectors are defined as $\bm{\delta}_1 = \mathbf{a}_1$, $\bm{\delta}_2 = \mathbf{a}_2$, and $\bm{\delta}_3 = -\mathbf{a}_1 - \mathbf{a}_2$, and $\phi_f$ is the phase of the nearest-neighbor hopping.
Additionally, we note that while the displacement field does not significantly affect the amplitude of the nearest-neighbor hopping, it can control its phase.
The value of $\phi_f$ changes from $\pi$ to $\sim 0.85\pi$ when the displacement field potential strength $\varepsilon_D$ is increased from $0\,\rm meV$ to $20\,\rm meV$.
Due to the ``band inversion'' around the $\gamma$ point, the hybridization between MLWF and the TPLO is not negligible.
Numerical calculation has also shown that the maximum value of $|V_{\rm hyb}(\vk)|$ in the MBZ can reach up to $7\sim 10 \rm \, meV$.
In Sec.~B of the SM \cite{supplemental_material}, we have provided a detailed discussion about the relevant numerical parameters of Eq.~(\ref{eqn:eff-kin}).

In Fig.~\ref{fig:disp-field-band}, we show the band structure and the orbital projections with displacement field potential strength $\varepsilon_D = 0\,\rm meV$ and $\varepsilon_D = 20\,\rm meV$.
It can be seen that with the displacement field potential up to $20 \rm\,meV$, the top most moir\'e band is predominantly contributed by the MLWF, with only very small contribution from the TPLO around the $\gamma$ point.
The hybridization between the MLWF and TPLO results in an avoided crossing around the $\gamma$ point, which is also evident in Fig.~\ref{fig:disp-field-band}.
Moreover, the real-space density distribution of the two orbitals under displacement field potential strengths $\varepsilon_D = 0\,\rm meV$ and $\varepsilon_D = 20\,\rm meV$ are shown in Figs.~\ref{fig:disp-field-band}(b,d).
We note that the density distribution of the two orbitals is not strongly dependent on the displacement field potential strength.

\begin{figure}
    \centering
    \includegraphics[width=1\linewidth]{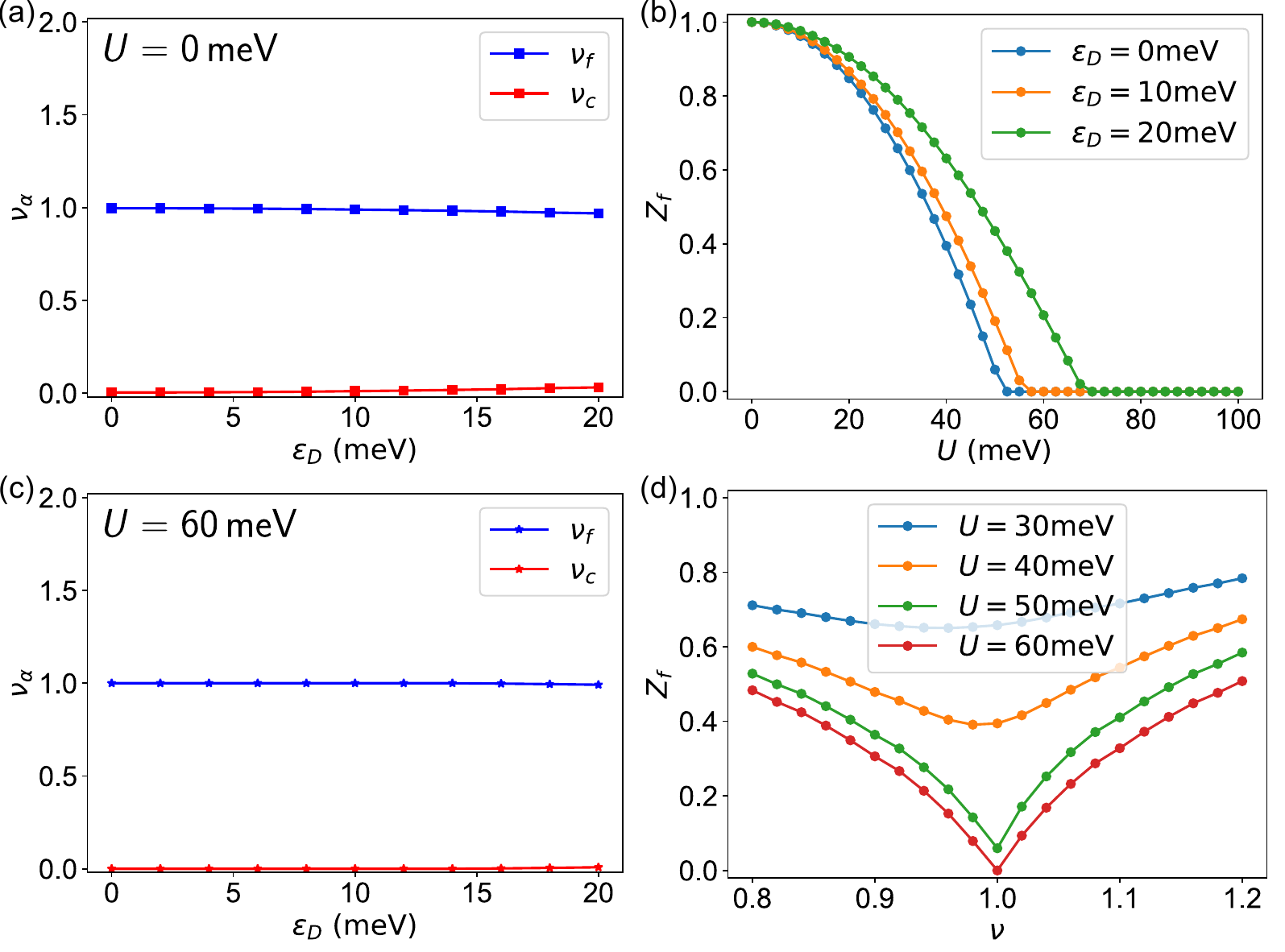}
    \caption{
    (a) The relative filling factors (hole picture) of the MLWF and the TPLO under different displacement field potential strengths, in the absence of interactions.
    (b) The quasiparticle weight of the MLWF as functions of interaction strength $0 \leq U \leq 100 \,\rm meV$, under different displacement field potential strength. 
    (c) The relative filling factors of the two orbitals under different displacement field potential strengths with an on-site interaction $U = 60\,\rm meV$.
    The total hole filling factor is fixed at $\nu = 1$.
    (d) The quasiparticle weight as a function of the hole filling factor $ 0.8 \leq \nu \leq 1.2$, for different interaction strength.
    The displacement field is fixed at $\varepsilon_D = 0\, \rm meV$.
    }
    \label{fig:interaction-scale}
\end{figure}

{\it \color{blue} Electronic correlations}---
Before we compute the interaction matrix elements in the MLWF and TPLO basis, we first analyze the relative filling factor of these two orbitals.
In the experiment \cite{Xia2024Unconventional}, the superconducting state is mostly observed when the top most band is nearly half-filled.
In Fig.~\ref{fig:interaction-scale}(a), we solve the relative (hole) filling factors of the MLWF and TPLO with total hole filling factor fixed at $\nu = 1$, without considering Coulomb interaction.
One can easily notice that the majority of holes accumulate in the MLFW, with only approximately $3\%$ occupying the TPLO when the displacement field potential strength is within the range $0\,{\rm meV}\leq \varepsilon_D \leq 20\,{\rm meV}$.
Since the MLWF is much closer to its half-filling, it is expected to exhibit significantly stronger correlation effects \cite{Kotliar1986, fazekas1999lecture, Hassan2010slave}.

As such, the minimum model which faithfully describes the low-energy effective physics of this system is given by the following Hamiltonian:
\begin{align}
    H_{\rm eff} &= H_0 + H_1\,,\label{eqn:eff-ham}\\
    H_1 &= U\sum_{\mathbf{R}}f^\dagger_{\mathbf{R}\uparrow}f_{\mathbf{R}\uparrow} f^\dagger_{\mathbf{R}\downarrow}f_{\mathbf{R}\downarrow}\,,
\end{align}
in which $H_0$ is defined in Eq.~(\ref{eqn:eff-kin}), and $H_1$ is the on-site Hubbard interaction for the MLWFs.
The value of interaction strength can be computed through the screened Coulomb potential and the wave function of the MLWF.
Numerical calculation under different displacement field demonstrates that the on-site interaction strength $\varepsilon U$ is about $550 \rm\,meV$ as discussed in detail in Sec.~C of the SM \cite{supplemental_material}.
In addition, it is not very sensitive to the displacement field.
Considering the fact that the dielectric constant of the hBN substrate is about $6$, and the dielectric constant of the single-layer WSe$_2$ is about $16$ \cite{Laturia2018}, we can estimate that the on-site Hubbard interaction strength $U$ is about $30 \sim 90 \rm\,meV$.
We also note that this estimation can be affected by the distance between the top and bottom gates, and it should be taken at the order-of-magnitude level instead of as a first-principle calculation.

Based on the parameter estimations above, we are now in position to address the correlation effects. 
We do so using self-consistent $U(1)$ slave-spin approach \cite{Yu2012U1slave}.
This method is able to estimate the suppression of quasiparticle weight under electron-electron correlation.
The technical detail of this approach is outlined in Sec.~D in the SM \cite{supplemental_material}.
To gain an overall perspective, we first perform calculations by varying the on-site interaction $U$.
With the total hole filling factor fixed at $\nu = 1$, we consider different displacement field potential strengths. 
The results can be found in Fig.~\ref{fig:interaction-scale}(b), in which the quasiparticle weight of the MLWF as the function of interaction strength $U$ is computed.
One can notice that an on-site interaction $U\sim 70 \,\rm meV$, which is somewhat larger than the topmost band's width of $\sim 40\,\rm meV$, is already strong enough to drive an orbital-selective Mott transition.
As estimated in Sec.~C in SM \cite{supplemental_material}, a reasonable estimation for the value of $U$ can reach up to $\sim 90\rm \,meV$, which is larger than the critical interaction strength predicted by the slave-spin approach.
Additionally, the presence of a non-vanishing displacement field can slightly reduce the electronic correlation for a fixed value of $U$.
These results suggest that the system is in the strongly correlated regime, where Landau quasiparticles are on the verge of being destroyed.

We next fix the value of interaction strength at $U = 60\,\rm meV$, and change the displacement field potential strength from $\varepsilon_D = 0\,\rm meV$ to $\varepsilon_D = 20\,\rm meV$.
The relative filling factors of the two orbitals are presented in Fig.~\ref{fig:interaction-scale}(c).
In comparison with Fig.~\ref{fig:interaction-scale}(b), the quasiparticle weight of the MLWF will increase from zero to a finite value with increasing $\varepsilon_D$.
Throughout this process, the majority of active degrees of freedom still originate from the MLWF rather than the TPLO, similar to the non-interacting case shown in Fig.~\ref{fig:interaction-scale}(a). 
This further justifies our effective model in Eq.~(\ref{eqn:eff-ham}) for capturing the low-energy correlation physics.

Finally, we also perform the simulation when the systems is doped away from $\nu = 1$, with 
interaction strength up to $U = 60\,\rm meV$ and displacement field potential strength set to $\varepsilon_D = 0\,\rm meV$. 
As shown in Fig.~\ref{fig:interaction-scale}(d), the quasiparticle weight of the MLWF is suppressed the most near the top band half filling point.

We expect that the quasiparticle weight of the MLWF at hole doping level $\nu - 1 \approx 0.05$ can be reduced to $\lesssim 0.2$.
This allows for an estimate \cite{xie2024kondo} of the Kondo ``coherence" temperature $T_{\rm coh} \sim \frac92 Z |t_f| \sim \rm 30\, K$.
Here the factor of $9/2$ comes from the ratio of the half band width of triangle lattice to the nearest-neighbor hopping.
The estimated coherent temperature is consistent with the value experimentally observed in Ref.~\cite{Xia2024Unconventional} (called $T^*$ there) at the order-of-magnitude level.

{\it \color{blue} Discussion}---
Several remarks are in order. 
First, related correlation physics arises in bulk materials with active flat bands. 
These include kagome and pyrochlore metals whose bare flat bands lie near the Fermi energy \cite{hu2023coupled,chen2024emergent, chen2023metallic,Huang2023np}, for which the construction of the compact molecular orbitals is also vitally important \cite{Souza2025,Mahankali2025}.
We also note that the effective interacting model that appears in the present work, which couples local degrees of freedom to extended orbitals that form a topological band, connects with the models and materials for Kondo-based metallic topology (Weyl-Kondo semimetals)~\cite{lai2018weyl,dzsaber2021giant}. 
As such, our work reveals new connections in the correlation physics among the different materials platforms.
More generally, we expect our work to crosstalk with the physics of other flat-band settings \cite{flat-perspective2024}, including moir\'{e} graphene systems \cite{Ram2021,Song2022,Kumar2022}.

Second, unlike the MLWF, whose gauge choice can be fixed by minimizing the real-space spread of its Wannier function, the gauge choice of the TPLO cannot be uniquely determined by simply finding its ``optimal'' Wannier function.
Instead, an extra gauge choice freedom, which is the position of the vortex singularity in the MBZ, remains \cite{Xie2024Chern}.
Different choice of vortex position will not affect the magnitude of the hybridization between the TPLO and MLWF, but it will affect the phase of this hybridization as well as the projected interactions in the TPLO.
The low occupancy of the TPLO for the total filling near $\nu=1$ justifies neglecting its interactions in our analysis.

Third, with a controlled basis for analyzing pairing tendencies, our work sets the stage to address the nature of the superconductivity state in tWSe$_2$ when the combination of active bands remains topological \cite{tj-2025}.

{\it \color{blue} Summary}---
We have studied the electronic structure of the tWSe$_2$ in the parameter regime with non-vanishing valley Chern numbers in the top two moir\'e bands.
We found that a compact molecular orbital, which predominantly describes the topmost moir\'e band, hybridizing with a topological power-law orbital associated with a Chern number $\mathcal{C} = -2$, which is far away from half-filling, can faithfully capture the low-energy space of these topological bands.
Based on this construction, we analyzed the strength of the electronic correlation effect in the localized orbital, which leads to a Kondo (Fermi) temperature scale that is consistent with the Kondo-lattice phenomenology observed in transport experiments.
Our construction provides a foundation for further understanding the unconventional pairings of its superconducting phase in the correlated topological band regime.
We expect that our work can be generalized to elucidate the correlation physics of a broad range of other systems with topologically obstructed active degrees of freedom.

\begin{acknowledgments}
    {\it Acknowledgments.~}We thank Lei Chen, Yuan Fang, Kin Fai Mak, Andrew Millis, Silke Paschen, Abhay Pasupathy, Jie Shan, Shouvik Sur, Yonglong Xie and Ming Yi for useful discussions. This work has been supported in part by the NSF Grant No.\ DMR-2220603 (F.X.), the AFOSR under Grant No.\ FA9550-21-1-0356 (C.L.), the Robert A. Welch Foundation Grant No.\ C-1411 (Q.S.) and the Vannevar Bush Faculty Fellowship ONR-VB N00014-23-1-2870 (Q.S.).
    J.C. acknowledges the support of the National Science Foundation under Grant No. DMR-1942447, support from the Alfred P. Sloan Foundation through a Sloan Research Fellowship and the support of the Flatiron Institute, a division of the Simons Foundation.
    The majority of the computational calculations have been performed on the Shared University Grid at Rice funded by NSF under Grant No.~EIA-0216467, a partnership between Rice University, Sun Microsystems, and Sigma Solutions, Inc., the Big-Data Private-Cloud Research Cyberinfrastructure MRI-award funded by NSF under Grant No. CNS-1338099, and the Extreme Science and Engineering Discovery Environment (XSEDE) by NSF under Grant No. DMR170109. 
    Q.S. acknowledges the hospitality of the Aspen Center for Physics, which is supported by NSF grant No. PHY-2210452.
\end{acknowledgments}

\bibliography{reference.bib}
\bibliographystyle{apsrev4-2}

\onecolumngrid
\clearpage

\beginsupplement
\section*{Supplemental Material}
\setcounter{secnumdepth}{3}

\tableofcontents

\section{Continuum model}

\begin{figure}
    \centering
    \includegraphics[width=0.75\linewidth]{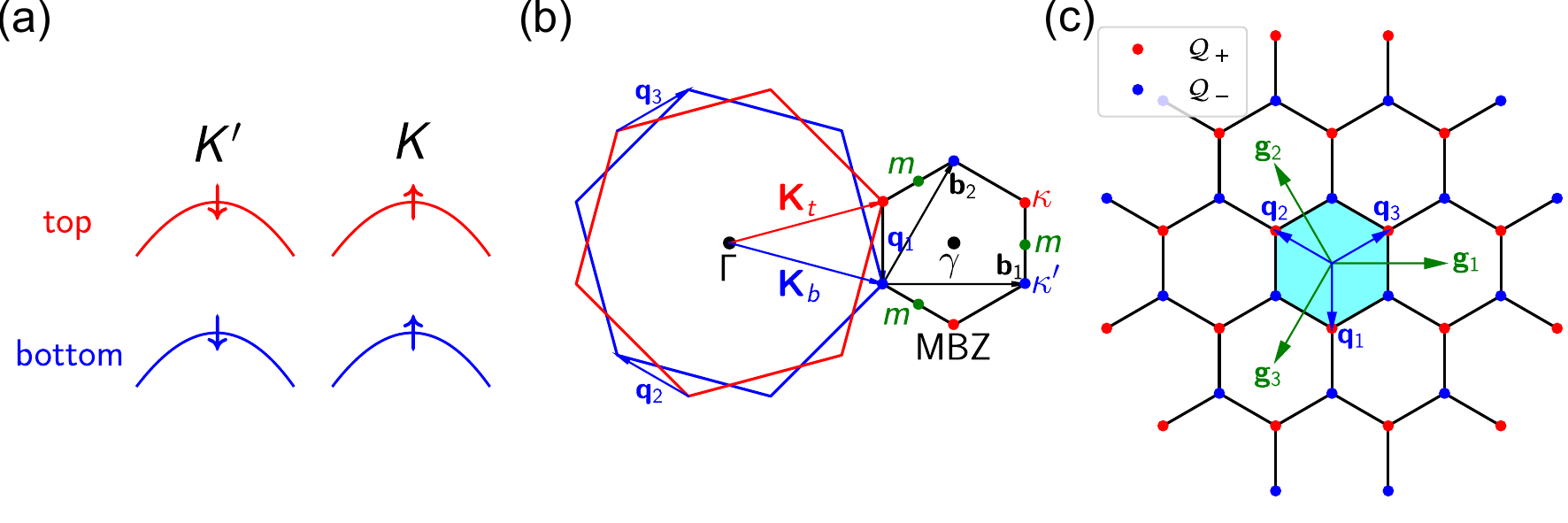}
    \caption{(a) Due to the strong spin-orbit coupling, the spin and valley degrees of freedom are locked to each other in single-layer TMDC materials.
    In twisted bilayer TMDC systems, the 
    (b) The moir\'e Brillouin zone is given by the hexagon with the red dashed line, and the triangular lattice is given by the hexagon with the blue dashed line. The reciprocal lattice vectors of the moir\'e Brillouin zone are labeled as $\mathbf{b}_1$ and $\mathbf{b}_2$. 
    The high-symmetry points are labeled.
    (c) The momentum space $\mathbf{Q}$ grids are spanned by recursively adding the vectors $\vq_{1,2,3}$. 
    Red and blue dots stand for the $K$ points from the top and bottom layers, respectively.
    The reciprocal vectors $\mathbf{g}_{1,2,3}$ are also labeled.
    }
    \label{fig:model-app}
\end{figure}

In this section, we briefly review the continuum model, which describes the single-valley band structure of twisted bilayer TMDC.
It is already well-known that the single-layer TMDC materials have a strong spin-orbit coupling, which locks the spin and valley degrees of freedom together \cite{Liu2013Three,Kormanyos2015kp}.
The low-energy effective theory of such materials is described by a quadratic hole band near the single layer $K$ and $K'$ points, as sketched in Fig.~\ref{fig:model-app}(a).
When the two layers are stacked on top of each other, electronic states with the same spin orientation will hybridize with each other, and a small twisting angle will lead to a moir\'e superlattice, corresponding to the moir\'e Brillouin zone (MBZ), which is presented in Fig.~\ref{fig:model-app}(b).
The single-valley effective continuum model for twisted bilayer TMDC materials can be written in the following form \cite{Wu2019topological,Devakul2021Magic}:
\begin{equation}
    h_0(\mathbf{r}) = \left(
        \begin{array}{cc}
            \frac{\nabla^2}{2m^*} + \tilde{v}_+(\mathbf{r}) + \frac{\varepsilon_D}{2}  & T(\mathbf{r}) \\
            T^*(\mathbf{r}) & \frac{\nabla^2}{2m^*} + \tilde{v}_-(\mathbf{r}) - \frac{\varepsilon_D}{2}
        \end{array}
    \right)\,,
\end{equation}
in which the two entries of the matrix stand for the top and bottom layers, respectively.
$\varepsilon_D$ denotes the potential difference between the two layers induced by a vertical displacement field, and the parameter $m^*$ is the effective mass of the hole pocket near the $K$ and $K'$ points of the single layer Brillouin zone. 
The intra-layer and inter-layer potentials are given by:
\begin{align}
    \tilde{v}_\ell(\mathbf{r}) =& 2\tilde{v} \sum_{j=1}^3 \cos\left(\mathbf{g}_j \cdot \mathbf{r} + \ell \psi\right)\,,\\
    T(\mathbf{r}) =& w \sum_{j=1}^3 e^{i\vq_j \cdot \mathbf{r}}\,.
\end{align}
Here the vectors $\mathbf{q}_{1,2,3}$ are the momentum difference between the $K$ points from the top and bottom layers,
and $\mathbf{g}_{1,2,3}$ are reciprocal vectors of the moir\'e superlattice, which are labeled in Fig.~\ref{fig:model-app}(c).
$\ell = \pm1$ stands for the top and bottom layers, respectively.
The values of the model parameters depend on the type of the TMDC materials. 
In twisted bilayer WSe$_2$, the effective mass is $m^* \approx 0.43 m_e$, the strength of the intra-layer potential $\tilde{v}$ is about $9 \rm\,meV$, the phase angle $\psi = 128^\circ$, and the inter-layer hopping amplitude $w$ is about $18 \rm\,meV$ \cite{Devakul2021Magic}. 

The continuum Hamiltonian is usually studied using the plane-wave basis.
For a plane-wave state from layer $\ell$, it can always be written as:
\begin{equation}
    \langle \mathbf{r}, \ell|\vk, \mathbf{Q} \rangle = \frac{1}{\sqrt{N\Omega_{c}}} e^{i(\vk - \mathbf{Q})\cdot \mathbf{r}}\,, ~~~ \mathbf{Q} \in \mathcal{Q}_\ell\,,
\end{equation}
where $N$ is the number of moir\'e unit cells, and $\Omega_c$ is the area of the moir\'e unit cell.
Using these plane wave basis, the matrix elements of the continuum Hamiltonian can be written as:
\begin{align}
    h_{\mathbf{Q}\mathbf{Q}'}(\vk) =& \frac{1}{N\Omega_c} \int d^2r \,e^{-i(\vk - \mathbf{Q}) \cdot \mathbf{r}}\left[h_0(\mathbf{r}) \right]_{\ell_\mathbf{Q}\ell_{\mathbf{Q}'}} e^{i(\vk - \mathbf{Q}')\cdot \mathbf{r}} \nonumber \\
    =& \left(-\frac{(\vk - \mathbf{Q})^2}{2m^*} + \ell_{\mathbf{Q}}\frac{\varepsilon_D}{2}\right)\delta_{\mathbf{Q}\mathbf{Q}'} + \tilde{v}\sum_{j = 1}^{3} \left(e^{i\psi} \delta_{\mathbf{Q} - \mathbf{Q}', \mathbf{g}_j} + e^{-i\psi} \delta_{\mathbf{Q} - \mathbf{Q}', -\mathbf{g}_j}\right) + w \sum_{j=1}^3 \left(\delta_{\mathbf{Q} - \mathbf{Q}', \vq_j} +  \delta_{\mathbf{Q} - \mathbf{Q}', -\vq_j}\right)\,,
\end{align}
in which $\ell_\mathbf{Q} = \pm 1$ if $\mathbf{Q} \in \mathcal{Q}_\pm$ is the layer index of the momentum lattice point $\mathbf{Q}$.
Diagonalizing the above Hamiltonian yields the Bloch states and the band structure of the twisted bilayer TMDC:
\begin{align}
    &\sum_{\mathbf{Q}'}h_{\mathbf{Q}\mathbf{Q}'}(\vk) u_{\mathbf{Q}',n}(\vk) = \varepsilon_n(\vk) u_{\mathbf{Q},n}(\vk)\,,\\
    &|\psi_{n\vk}\rangle = \sum_{\mathbf{Q}\in\mathcal{Q}_\pm} u_{\mathbf{Q},n}(\vk) |\vk, \mathbf{Q}\rangle\,.
\end{align}
Using the parameters of twisted bilayer WSe$_2$ with a twisting angle $\theta = 3.65^\circ$ and displacement field potential strength $\varepsilon_D = 0 \,\rm meV$, we can compute the single-valley band structure, and it has been shown in Fig.~1(b) in the main text.

In addition, the symmetries can also be analyzed easily in the plane wave basis.
For example, the spinful $C_{3z}$ symmetry can be represented by the following unitary matrix in the plane wave basis:
\begin{equation}
    [D(C_{3z})]_{\mathbf{Q}\mathbf{Q}'} = e^{i\frac{\pi}{3}}\delta_{\mathbf{Q},C_{3z}\mathbf{Q}'}\,,
\end{equation}
in which the phase factor comes from the spin rotation.
Using this representation matrix, we can also compute the rotation eigenvalues of the Bloch bands at high symmetry points $\gamma$, $\kappa$ and $\kappa'$:
\begin{equation}\label{eqn:C3-eig}
    \xi_\mathbf{K}(C_{3z}) = \sum_{\mathbf{Q} \mathbf{Q}'} u^*_{\mathbf{Q},n}(C_{3z}\mathbf{K})[D(C_{3z})]_{\mathbf{Q}\mathbf{Q}'} u_{\mathbf{Q}', n}(\mathbf{K})\,.
\end{equation}
The $C_{3z}$ eigenvalues of the top two moir\'e bands at these high symmetry points labeled in Fig.~1(b) in the main text are computed via Eq.~(\ref{eqn:C3-eig}).

\section{Construction of the orbitals}

In this section, we discuss the technical detail of the ``partial Wannierization'' procedure.
We project the Bloch states of the top two bands onto a trial Gaussian orbital centered at the $1a$ Wyckoff position on a $24 \times 24$ momentum grid and provide this data to the \textsc{Wannier90} \cite{Marzari1997Maximally,Souza2001Maximally,Pizzi2020wannier} software as its input, and it returns the rectangular ``disentanglement'' matrix $U^{\rm dis}_{n,f}(\vk)$ as the output, which ``rotates'' the top two bands' Bloch states into the Bloch wave function of the MLWF:
\begin{equation}
    \tilde{u}_{\mathbf{Q},f}(\vk) = \sum_{n = 1}^2 u_{\mathbf{Q}, n}(\vk)U^{\rm dis}_{n, f}(\vk)\,.
\end{equation}
The wave function of the other orbital, denoted as $\tilde{u}_{\mathbf{Q}, c}(\vk)$, can be constructed by an orthonormalization procedure:
\begin{equation}
    \scalemath{0.9}{\tilde{u}_{\mathbf{Q},c}(\vk) \tilde{u}^{*}_{\mathbf{Q}',c}(\vk) = \sum_{n = 1}^2 u_{\mathbf{Q}, n}(\vk) u^{*}_{\mathbf{Q}',n}(\vk) - \tilde{u}_{\mathbf{Q},f}(\vk) \tilde{u}^{*}_{\mathbf{Q}',f}(\vk) \,.}
\end{equation}
However, the wave function constructed from the above projection opeartor still needs gauge fixing, as the phase factors of $\tilde{u}_{\mathbf{Q},c}(\vk)$ at different $\vk$ are not determined.
Since the top two bands carry a total Chern number $\mathcal{C} = -2$, and the MLWF orbital is already an exponentially localized Wannier function, the above wave function described by $\tilde{u}_{\mathbf{Q},c}(\vk)$ will carry a Chern number $\mathcal{C} = -2$.
Therefore, we can fix the gauge of $\tilde{u}_{\mathbf{Q},c}(\vk)$ using the algorithm described in Ref.~\cite{Xie2024Chern}, placing a vortex with a vorticity of $-2$ at the $\gamma$ point in the moir\'e Brillouin zone, which will lead to a power-law decaying Wannier function \cite{thouless_wannier_1984}.
The Wannier function of both orbitals can be computed via Fourier transformation of these gauge-fixed Bloch states:
\begin{equation}\label{eqn:wannier-real-space}
    \tilde{W}_\alpha(\mathbf{r}, \ell) = \frac{1}{N\sqrt{\Omega_c}}\sum_{\vk \in {\rm MBZ}} \sum_{\mathbf{Q} \in \mathcal{Q}_\ell} \tilde{u}_{\mathbf{Q},\alpha}(\vk)e^{i(\vk - \mathbf{Q})\cdot \mathbf{r}}\,.
\end{equation}
The real-space density distribution plots in Figs.~2(b,d) in the main text are computed using Eq.~(\ref{eqn:wannier-real-space}).

We can then project the continuum Hamiltonian into the Hilbert space spanned by the above two orbitals. 
The matrix elements can be written as:
\begin{align}
    \varepsilon_f(\vk) &= \sum_{\mathbf{Q}\mathbf{Q}'} \tilde{u}^*_{\mathbf{Q},f} h_{\mathbf{Q}\mathbf{Q}'}(\vk) \tilde{u}_{\mathbf{Q}',f}(\vk)\,,\\
    \varepsilon_c(\vk) &= \sum_{\mathbf{Q}\mathbf{Q}'} \tilde{u}^*_{\mathbf{Q},c}(\vk) h_{\mathbf{Q}\mathbf{Q}'}(\vk) \tilde{u}_{\mathbf{Q}',c}(\vk)\,,\\
    V_{\rm hyb}(\vk) &= \sum_{\mathbf{Q}\mathbf{Q}'} \tilde{u}^*_{\mathbf{Q},f}(\vk) h_{\mathbf{Q}\mathbf{Q}'}(\vk) \tilde{u}_{\mathbf{Q}',c}(\vk)\,,
\end{align}
which all can be evaluated numerically.
Hopping amplitudes among the Wannier states can be solved from these functions through an inverse Fourier transformation.

\begin{figure}
    \centering
    \includegraphics[width=0.7\linewidth]{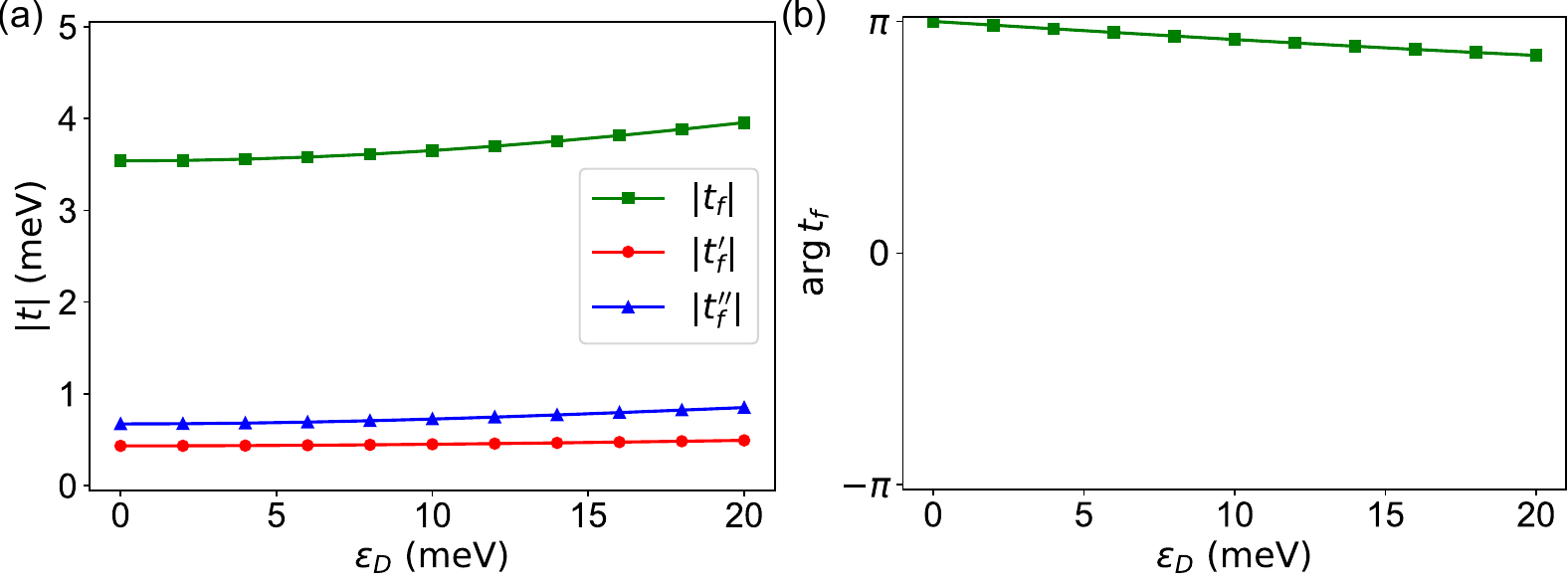}
    \caption{(a) The absolute value of the nearest-neighbor ($t_f$), next-nearest-neighbor ($t_f'$) and next-next-nearest-neighbor ($t_f''$) hopping amplitudes among the MLWFs. (b) The phase angle of the nearest-neighbor hopping amplitude $t_f$ along the $\mathbf{a}_1$ direction.}
    \label{fig:disp-hopping}
\end{figure}

In Fig.~\ref{fig:disp-hopping}(a), we show the magnitude of the nearest-neighbor hopping $t_f$, next-nearest-neighbor hopping $t_f'$ and next-next-nearest-neighbor hopping $t_f''$ under different displacement field potential strengths ($0 \leq \varepsilon_D \leq 20 {\rm \, meV}$).
We note that, within a reasonably large interval of displacement field potential, the hopping amplitudes between the MLWFs are dominated by the nearest-neighbor hopping.
The absolute value of this hopping amplitude is about $3.5 \sim 4.0 \rm\,meV$, and its phase angle is also shown in Fig.~\ref{fig:disp-hopping}(b).
At zero displacement field, this hopping amplitude is real and negative.
With the increase of the displacement field, the absolute value of the hopping amplitude slightly increases, and the phase angle also slightly deviates from $\pi$.

The discussions in previous paragraphs about the continuum model and the partial Wannierization procedure are mostly based on the single-valley ($K$-valley, spin $\uparrow$) Hamiltonian. 
The corresponding wave functions from the opposite spin-valley sector ($K'$-valley, spin $\downarrow$) can be obtained by the time-reversal operation, which flips the momentum $\vk, \mathbf{Q}$ and takes the complex conjugate of the wave function: 
\begin{align}
    \tilde{u}^{(\uparrow)}_{\mathbf{Q},\alpha}(\vk) &= \tilde{u}_{\mathbf{Q}, \alpha}(\vk) \,,\\
    \tilde{u}^{(\downarrow)}_{\mathbf{Q},\alpha}(\vk) &= \tilde{u}^*_{-\mathbf{Q}, \alpha}(-\vk)\,,
\end{align}
in which $\alpha=c,f$ stands for the orbital indices.
Matrix elements projected into the $K'$-valley spin $\downarrow$ Bloch states can also be obtained accordingly.
As a summary, the top two bands of the continuum model can be well described by the following effective Hamiltonian:
\begin{equation}\label{eqn:kin-ham-app}
    H_0 = \sum_{\vk,\sigma} \varepsilon_f^{(\sigma)}(\vk) f^\dagger_{\vk\sigma}f_{\vk\sigma} + \sum_{\vk, \sigma} \varepsilon^{(\sigma)}_c(\vk) c^\dagger_{\vk\sigma} c_{\vk\sigma} + \sum_{\vk, \sigma} \left(V^{(\sigma)}_{\rm hyb}(\vk) f^\dagger_{\vk\sigma} c_{\vk\sigma} + {\rm h.c.}\right)\,.
\end{equation}
The dispersion of the MLWF can also be well-approximated by a nearest-neighbor hopping model:
\begin{equation}
    \varepsilon_f^{(\sigma)}(\vk) \approx 2|t_f| \sum_{i = 1}^{3}\cos(\vk \cdot \bm{\delta}_i + \sigma \phi_f) + \tilde{\epsilon}_f\,,
\end{equation}
where $\tilde{\epsilon}_f$ is the ``on-site potential'' of the MLWF, and the Bravais lattice vectors are defined as $\bm{\delta}_1 = \mathbf{a}_1$, $\bm{\delta}_2 = \mathbf{a}_2$, and $\bm{\delta}_3 = -\mathbf{a}_1 - \mathbf{a}_2$.
The conduction electron dispersions in the two spin sectors are given by $\varepsilon^{(\uparrow)}_c(\vk) = \varepsilon_c(\vk)$ and $\varepsilon^{(\downarrow)}_c(\vk) = \varepsilon_c(-\vk)$, and the hybridization functions are given by $V^{(\uparrow)}_{\rm hyb}(\vk) = V_{\rm hyb}(\vk)$ and $V^{(\downarrow)}_{\rm hyb}(\vk) = V^*_{\rm hyb}(-\vk)$.
Note that this is {\it not} a tight-binding model in the conventional sense, as the ``conduction electrons'' $c,c^\dagger$ are not degrees of freedom associated with localized orbitals.

\begin{figure}
    \centering
    \includegraphics[width=\linewidth]{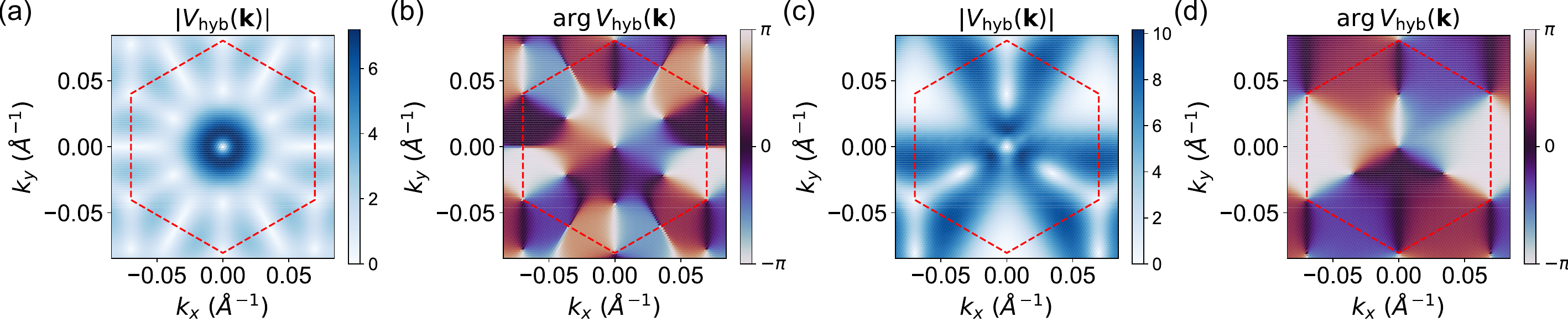}
    \caption{(a) The absolute value of the hybridization function $|V_{\rm hyb}(\vk)|$ over the MBZ with displacement field potential strength $\varepsilon_D = 0\,\rm meV$.
    The red dashed line stands for the MBZ.
    (b) The phase of $V_{\rm hyb}(\vk)$ over the MBZ. 
    Note there are multiple vortices in the MBZ.
    (c-d) The absolute value and the phase of the hybridization function with diplacement field potential strength $\varepsilon_D = 20\,\rm meV$.
    }
    \label{fig:hyb}
\end{figure}

The blue and red dashed lines in Figs.~2(b,d) in the main text are the ``dispersion relationships'' $\varepsilon_c(\vk)$ and $\varepsilon_f(\vk)$ without the hybridization terms.
Additionally, the function $V_{\rm hyb}(\vk)$ in the first Brillouin zone with displacement field potential strengths $\varepsilon_D = 0\,\rm meV$ and $\varepsilon_D = 20\,\rm meV$ can also be found in Fig.~\ref{fig:hyb}.
The maximal absolute value of the hybridization $|V_{\rm hyb}(\vk)|$ is around $7\sim 10\,\rm meV$.

\section{Coulomb interaction strength}

\begin{figure}
    \centering
    \includegraphics[width=0.5\linewidth]{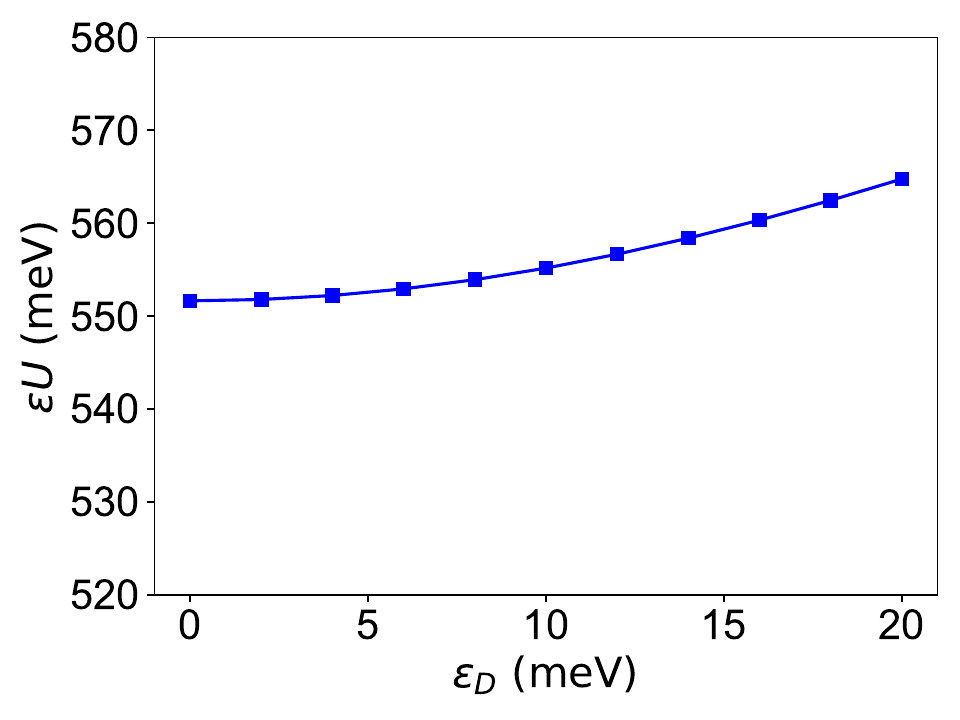}
    \caption{The on-site Hubbard interaction strength $U$ in the MLWF basis, as a function of the displacement field potential $\varepsilon_D$.
    We note that the value of $\varepsilon U$ is about $550 \rm\,meV$ and it is not very sensitive to the displacement field.
    Here we assumed the gate distance is $\xi = 10\,\rm nm$, and the MBZ is discretized by a $6 \times 6$ grid.}
    \label{fig:disp-U}
\end{figure}

In this section, we study the strength of the projected Coulomb interactions in the MLWF constructed in the previous section.
In the Hilbert space spanned by the above two orbitals, the projected Coulomb interaction can be written as:
\begin{equation}
    H_{\rm int} = \frac{1}{2N \Omega_{c}} \sum_{\vq\vk\vk'}\sum_{\sigma\sigma'}\sum_{\alpha\beta\alpha'\beta'} \tilde{\mathcal{U}}_{\alpha\beta;\alpha'\beta'}^{(\sigma\sigma')} (\vq;\vk,\vk') \alpha^\dagger_{\vk+\vq,\sigma} \beta_{\vk,\sigma} \alpha'^\dagger_{\vk'-\vq,\sigma'} \beta'_{\vk',\sigma'}\,, 
\end{equation}
in which the fermion operators $\alpha, \beta, \alpha', \beta'$ can be either $c$ or $f$.
The matrix elements in the interacting Hamiltonian can be computed using the Bloch wave functions of the continuum model:
\begin{equation}
    \tilde{\mathcal{U}}_{\alpha\beta;\alpha'\beta'}^{(\sigma\sigma')}(\vq;\vk,\vk') = \sum_{\mathbf{G}}\mathcal{V}(\vq + \mathbf{G}) \sum_{\mathbf{Q}}\tilde{u}^{(\sigma)*}_{\mathbf{Q},\alpha}(\vk + \vq + \mathbf{G}) \tilde{u}^{(\sigma)}_{\mathbf{Q},\beta}(\vk) \sum_{\mathbf{Q}'} \tilde{u}^{(\sigma')*}_{\mathbf{Q}',\alpha'}(\vk' - \vq - \mathbf{G}) \tilde{u}^{(\sigma')}_{\mathbf{Q}',\beta'}(\vk')\,,
\end{equation}
where $\mathbf{G}$ stands for all reciprocal vectors of the moir\'e superlattice.
$\mathcal{V}(\vq) = (\xi e^2/4\varepsilon_0\varepsilon) \tanh(\xi q /2)/(\xi q/2)$ is the Fourier transformation of the screened Coulomb potential, $\xi$ is the distance between the two metallic gates, and $\varepsilon$ is the dielectric coefficient of the substrate.
Fourier transforming this interacting Hamiltonian into the Wannier basis, we have:
\begin{align}
    &H_{\rm int} = \frac{1}{2} \sum_{\mathbf{R}_0}\sum_{\mathbf{R}\mathbf{d}\mathbf{d}'}\sum_{\sigma\sigma'}\sum_{\alpha\beta\alpha'\beta'} \mathcal{U}^{(\sigma\sigma')}_{\alpha\beta;\alpha'\beta'}(\mathbf{R};\mathbf{d},\mathbf{d} ') \alpha^\dagger_{\mathbf{R}+\mathbf{d}+\mathbf{R}_0,\sigma} \beta_{\mathbf{R} + \mathbf{R}_0,\sigma} \alpha'^\dagger_{\mathbf{d}' + \mathbf{R}_0, \sigma'} \beta'_{\mathbf{R}_0,\sigma'} \,,\\
    &\mathcal{U}^{(\sigma\sigma')}_{\alpha\beta;\alpha'\beta'}(\mathbf{R}; \mathbf{d},\mathbf{d}') = \frac{1}{N^3\Omega_c} \sum_{\vq\vk\vk'}e^{i\vq\cdot(\mathbf{R} + \mathbf{d} - \mathbf{d}')} e^{i\vk\cdot \mathbf{d}}e^{i\vk'\cdot \mathbf{d}'}\tilde{\mathcal{U}}_{\alpha\beta;\alpha'\beta'}^{(\sigma\sigma')}(\vq;\vk,\vk')\,.
\end{align}
Due to the exponentially localized nature of the MLWF, the interaction matrix elements will also be dominated by the on-site Hubbard interactions. 
The value of interaction strength can be computed through the following equation:
\begin{equation}
    U = \mathcal{U}^{(\uparrow\downarrow)}_{ff;ff}(\mathbf{0};\mathbf{0},\mathbf{0}) = \frac{1}{N^3 \Omega_c}\sum_{\vq\vk\vk'}\tilde{\mathcal{U}}^{(\uparrow\downarrow)}_{ff;ff}(\vq;\vk,\vk')\,.
\end{equation}
Numerical calculation under different displacement field demonstrates that the on-site interaction strength $\varepsilon U$ is about $550 \rm\,meV$, and it is not very sensitive to the displacement field, as shown Fig.~\ref{fig:disp-U}.
Considering the fact that the dielectric constant of the hBN substrate is about $6$, and the dielectric constant of the single-layer WSe$_2$ is about $16$ \cite{Laturia2018}, we can estimate that the on-site Hubbard interaction strength $U$ is about $30 \sim 90 \rm\,meV$.
We also note that this estimation can be affected by the distance between the top and bottom gates.
Thus, it represents an order-of-magnitude estimation, instead of a first-principle calculation.

\section{\texorpdfstring{$U(1)$}{U(1)}-slave spin method}\label{appsec:ss}

In this section, we briefly review the $U(1)$ slave spin approach, which is useful in qualitatively describing the correlation effects in strongly correlated electronic systems \cite{Yu2012U1slave}.
We consider a multi-band interacting Hamiltonian with the following form:
\begin{equation}\label{eqn:FullHam}
    H = H_0 + H_{1}\,,
\end{equation}
in which the kinetic Hamiltonian $H_0$ can be expressed in the form shown in Eq.~(\ref{eqn:kin-ham-app}), and the interaction Hamiltonian is given by a simple on-site Hubbard term:
\begin{align}\label{eqn:interaction}
    H_1 = U \sum_{\mathbf{R},\sigma} \tilde{n}_{f \mathbf{R}\uparrow} \tilde{n}_{f\mathbf{R}\downarrow} \,,
\end{align}
where $\tilde{n}_{f\mathbf{R}\sigma}=f^\dagger_{\mathbf{R}\sigma}f  _{\mathbf{R}\sigma} - \frac12$ is the relative fermion number operator of the MLWF.

In the framework of this $U(1)$ slave spin theory, a {\it local} fermionic operator for the MLWF $f^\dagger_{\mathbf{R}\sigma}$ is represented by the product of a spin-$\frac12$ bosonic operator $o^\dagger_{\mathbf{R}\sigma}$ (``slave spin'') and another fermionic operator $\chi^\dagger_{\mathbf{R}\sigma}$ (``slave fermion''):
\begin{equation}\label{eqn:parton}
    f^\dagger_{\mathbf{R}\sigma} \rightarrow \chi^\dagger_{\mathbf{R}\sigma}o^\dagger_{\mathbf{R}\sigma}\,,
\end{equation}
where the spin operator $o^\dagger$ has the following form:
\begin{equation}\label{eqn:slave_o}
    o^\dagger_{\mathbf{R}\sigma} = P^+_{\mathbf{R}\sigma} S^+_{\mathbf{R}\sigma}P^-_{\mathbf{R}\sigma}\,,~~~P^\pm_{\mathbf{R}\sigma} = \frac{1}{\sqrt{\frac{1}{2} \pm S^z_{\mathbf{R}\sigma}}}\,.
\end{equation}
We note that this construction enlarges the local Hilbert space dimension.
In order to guarantee that the solution is within the physical Hilbert space at the
saddle-point level, a Lagrange multiplier term has to be added into the parton Hamiltonian:
\begin{equation}\label{eqn:lagrange_mult}
    H_\lambda = \sum_{\mathbf{R}\sigma} \lambda\left(S^z_{\mathbf{R}\sigma} + \frac12 - \chi^\dagger_{\mathbf{R}\sigma} \chi_{\mathbf{R}\sigma}\right)\,.
\end{equation}
Hence, the local constraint $\langle S^z_{\mathbf{R}\sigma} \rangle + 1/2 = \langle \chi^\dagger_{\mathbf{R}\sigma} \chi_{\mathbf{R}\sigma} \rangle$ can be satisfied at the saddle-point level by considering $\lambda$ as another variational parameter. 
At the saddle-point level and based on a ``single-site approximation'' for the parton operators, the full interacting Hamiltonian can be decoupled into an interacting impurity slave-spin term $H^S$, and a ``non-interacting'' slave-fermion term $H^f$.
The slave-spin Hamiltonian takes the following form:
\begin{equation}
    H^S = U S^z_{\uparrow} S^z_{\downarrow} + \lambda \sum_{\sigma} S^z_{\sigma} + \sum_{ \sigma}\left[h_0 \frac{S^+_{\sigma}}{\sqrt{n_0(1 - n_0)}} + {\rm h.c.}\right]\,, 
\end{equation}
in which the coordinate index $\mathbf{R}$ for the slave-spin operators $S_{\sigma}^{z,\pm}$ is omitted due to the single-site approximation. 
The bath field $h_0$ is determined from the correlation functions of the slave-fermion operators:
\begin{equation}
    h_0 = \frac{1}{N}\sum_{\vk} V^{(\sigma)}_{\rm hyb}(\vk) \langle \chi^\dagger_{\vk\sigma} c_{\vk\sigma} \rangle + \frac{1}{N}\sum_{\vk} \sqrt{Z} \left(\varepsilon_f^{(\sigma)}(\vk) - \tilde{\epsilon}_f\right) \langle \chi^\dagger_{\vk\sigma} \chi_{\vk \sigma} \rangle\,. \\
\end{equation}
The quasiparticle weight of the MLWF can be determined from the solution of the slave-spin Hamiltonian:
\begin{equation}
    \sqrt{Z} = \langle o^\dagger_{\sigma} \rangle = \frac{\langle S^+_{\mathbf{\sigma}} \rangle}{\sqrt{n_f (1-n_f)}}\,,
\end{equation}
where $n_f = \langle \chi^\dagger_{\mathbf{R}\sigma} \chi_{\mathbf{R}\sigma} \rangle$ is the fermion density expectation value of the MLWF.
On the other side, the slave-fermion Hamiltonian takes the following form:
\begin{align}
    H^f =& \sum_{\vk,\sigma} Z\left(\varepsilon_f^{(\sigma)}(\vk) -\tilde{\epsilon}_f \right)\chi^\dagger_{\vk\sigma} \chi_{\vk\sigma} + \sum_{\vk,\sigma}\left(\tilde{\epsilon}_f - \lambda + \lambda_0 - E_F\right) \chi^\dagger_{\vk\sigma}\chi_{\vk\sigma} \nonumber \\
    & \sum_{\vk,\sigma} \left(\varepsilon_c^{(\sigma)}(\vk) - E_F\right) c^\dagger_{\vk\sigma}c_{\vk\sigma} + \sqrt{Z}\sum_{\vk,\sigma}\left(V_{\rm hyb}^{(\sigma)}(\vk) \chi^\dagger_{\vk\sigma}c_{\vk\sigma} + {\rm h.c.}\right)\,.
\end{align}
Here, the parameter $\lambda^0$ is given by the following expression:
\begin{equation}
    \lambda^0 = -\sqrt{Z}|h_0|\frac{2 n_f - 1}{n_f(1 - n_f)}\,, \\
\end{equation}
which guarantees the slave-fermion Hamiltonian reduces to $H_0$ in the $U = 0$ limit.

We note that the parameter $Z$ in $H^f$ is determined from the ground state of $H^S$, while the parameter $h_0$ in $H^S$ is determined from the ground state of $H^f$.
Hence, for a given total filling factor, all these parameters $\sqrt{Z}$, $\lambda$, $\lambda^0$ and the Fermi energy $E_F$ can be solved self-consistently together.

\end{document}